\documentclass[ACS,STIX1COL]{WileyNJD-v2}

\articletype{Article}%

\def\r{{\mathbf{r}}}

\received{<day> <Month>, <year>}
\revised{<day> <Month>, <year>}
\accepted{<day> <Month>, <year>}

\begin{document}

\title{{L}ike-charge attraction in one- and two-dimensional Coulomb systems}

\author[1]{Gabriel T\'ellez*}

\authormark{G. T\'ellez}

\address[1]{\orgdiv{Departamento de F\'isica}, \orgname{Universidad de los Andes}, \orgaddress{\state{Bogot\'a}, \country{Colombia}}}

\corres{*\email{gtellez@uniandes.edu.co}}

\abstract[Summary]{The bare Coulomb interaction between two like-charges is
repulsive. When these charges are immersed in an electrolyte, the thermal
fluctuations of the ions turn the bare Coulomb interaction into an effective
interaction between the two charges. An interesting question arises: is it
possible that the effective interaction becomes attractive for like-charges? We
will show how this like-charge attraction phenomenon is indeed predicted in some
one- and two-dimensional models of Coulomb systems. Exact analytical results can
be obtained for these Coulomb systems models due to some connections that they
have with integrable field theories. We will explain the techniques that allow
obtaining exact results for the correlation functions between like-charge
particles and how, under certain circumstances, the natural repulsion
interaction shifts to being attractive. Although the technical details differ
for 1d or 2d systems, the physical cause of this phenomenon is rooted in a
three-body interaction between the two like-charges and an ion of the
electrolyte with an opposite charge.  
}



\maketitle

\footnotetext{\textbf{Abbreviations:} 1d: one-dimensional, 2d: two-dimensional,
3d: three-dimensional}

\section{Introduction}
\label{sec:intro}

Like-charge attraction is the intriguing phenomenon when two particles with
charges of the same sign that are immersed in an electrolyte turn out to have an
effective interaction that is attractive instead of the natural expected
repulsion. The effective interaction accounts for all the collective effects of
the particles and the surrounding ions of the electrolyte that are subject to
thermal fluctuations. The like-charge attraction phenomenon has been evidenced
in several experiments~\cite{Ise83, Larsen97} and numerical
simulations~\cite{Jensen98, Linse99}. For some time a satisfactory explanation
of this phenomenon was lacking. It was shown that it is impossible to have
like-charge attraction in the low coulombic coupling regime, where the effective
interactions can be computed using Poisson-Boltzmann theory and are always
repulsive~\cite{Neu99, TrizacRaimbault99, Trizac2000}. A plausible explanation
to the experimental results from~\cite{Larsen97} was given in~\cite{Squires2000}
where a nonequilibrium hydrodynamic interaction could explain the like-charge
attraction. Nevertheless the question of whether the like-charge attraction
could be a consequence of only the electrostatic collective interactions
remained open for some time. Advances in understanding the strongly Coulomb
coupled regime~\cite{Rouzina96, Shklovskii02, Levin02, Netz01} shed some light
on that question. A systematic high coupling expansion~\cite{SamajTrizac11}
revealed that in the strongly coupled regime like-charges can have an attractive
interaction in some particular set of the parameters (coupling and distance
between the particles)~\cite{TrizacSamaj13}. 

In parallel, there have been many advances in the theory of exactly solvable
models of Coulomb systems in one
dimension~\cite{Lenard61,Lenard62,Prager62,Dean09} and two
dimensions~\cite{Janco81,Alastuey81,Gaudin85,Cornu87,Samaj2000}. These are
simplified models of point-charged particles living in $d=1$ or $d=2$ dimensions
and interacting with the Coulomb potential corresponding to that dimension,
namely, for two unit charges separated by a distance $r$,
\begin{equation}
    v_d(r) =
    \begin{cases}
        - r &\text{for }d=1 \\
        -\ln r &\text{for }d=2
    \end{cases}
    .
\end{equation}
These models provide a testing ground to understand several properties of classical
charged systems. They also exhibit interesting connections with integrable
quantum field theories. In this article, we present how the like-charge attraction
phenomenon is predicted in several solvable models in one and two dimensions. In
the section~\ref{sec:1d}, we review results regarding like-charge attraction in
1d models. In section~\ref{sec:2d}, we present some 2d models and extend some
recent results to study the possibility of like-charge attraction in
charge-asymmetric plasmas. Finally, we conclude with an overview of the
like-charge attraction in the context of solvable models.

\section{One-dimensional models}
\label{sec:1d}

Consider a system of two equal charges $Q$ located in a line ($x$-axis) at $x=0$
and $x=L$ and screened by $N$ counterions with charge $e=-2Q/N$. The system is
globally neutral. The charges $Q$ are at fixed positions, while the counterions
can move between the two charges $Q$ and are at thermal equilibrium at a
temperature $T$. As usual we define $\beta =1/(k_B T)$ with $k_B$ the Boltzmann
constant. In one dimension, the potential energy between two particles of
charges $q$ and $q'$ located at $x$ and $x'$ is 
\begin{equation}
    u_1(x,x')=-qq'|x-x'|
    \,.
\end{equation}
In one dimension, the charges $q$ have dimensions of
$(\text{energy/length})^{1/2}=\text{force}^{1/2}$. Therefore one can define a
characteristic length comparing the thermal energy with the electrostatic energy
\begin{equation}
    l_B^{1d} = k_B T /e^2
    \,.
\end{equation}
This is the one-dimensional equivalent of the Bjerrum length $l_B^{3d}$. Notice
however the difference with the 3d situation where $l_B^{3d}=e_{3d}^2/(k_B T)$
for $e_{3d}$ the elementary 3d charge. This difference is due to the change in
the Coulomb potential from $1/r$ in 3d to $-r$ in 1d. The 1d-charges have
dimensions of 3d-charges/distance$^2$. The total potential energy $U$ of the
system is given by
\begin{equation}
    \label{eq:U}
    \beta U = -\sum_{1\leq i < j \leq N} \frac{|x_i - x_j|}{l_B^{1d}} +
    \left(\frac{N}{2}\right)^2 \frac{L}{l_B^{1d}}
    \,.
\end{equation}

At this point is useful to recall some properties of 1d electrostatics. The
electric field at $x'$ created by a single charge $q$ located at $x$ is constant
and equal to $q$ for $x'>x$ (right side of $q$) and equal to $-q$ for $x'<x$
(left side). Therefore, the counterions feel a constant force so long they do
not interchange positions with their neighbors. To obtain the force on a given
counterion one simply has to sum all charges at its right and subtract the
charges at its left. This greatly simplifies the analysis of the system to the
point that by ordering the particle positions one can rewrite the potential
energy~(\ref{eq:U}) as a sum of differences of consecutive positions of the
particles ($x_k-x_{k+1}$) as shown in~\cite{TellezTrizac15}. With this one
recognizes that the partition function of the system is an $N$-fold convolution
product that can be explicitly computed using the Laplace transform which is
equivalent to work in the isobaric ensemble~\cite{TellezTrizac15}. From there
the effective force between the two charges $Q$ is obtained as the derivative of
the canonical partition function with respect to $L$. From those exact results,
the possibility of like-charge attraction appears when $N=2p+1$ is an odd integer
and for large distances $L$. 

\begin{figure}
    \begin{center}
        \includegraphics[width=0.8\textwidth]{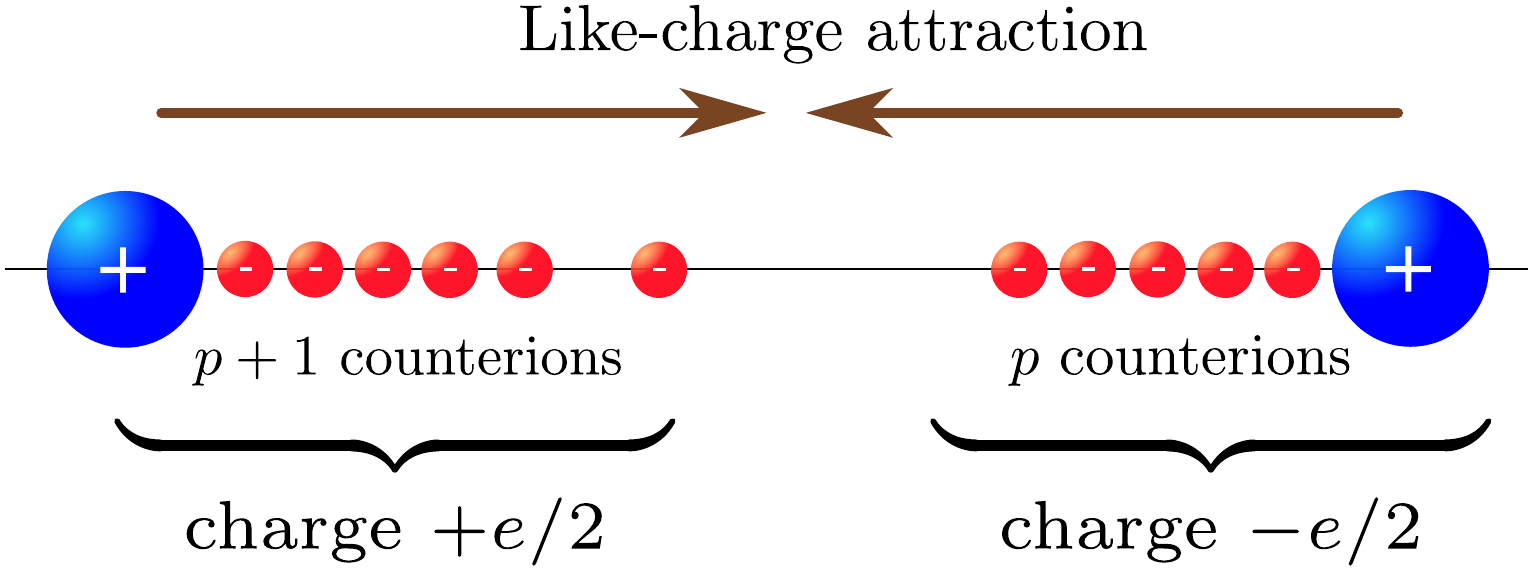}
    \end{center}
    \caption{
        \label{fig:attract1d}
        Illustration of the like-charge attraction in 1d systems due to an
        imbalance of charge when an odd number $N=2p+1$ of counterions is
        present. 
        }  
\end{figure}

\begin{figure}
    \begin{center}
        \includegraphics[width=0.8\textwidth]{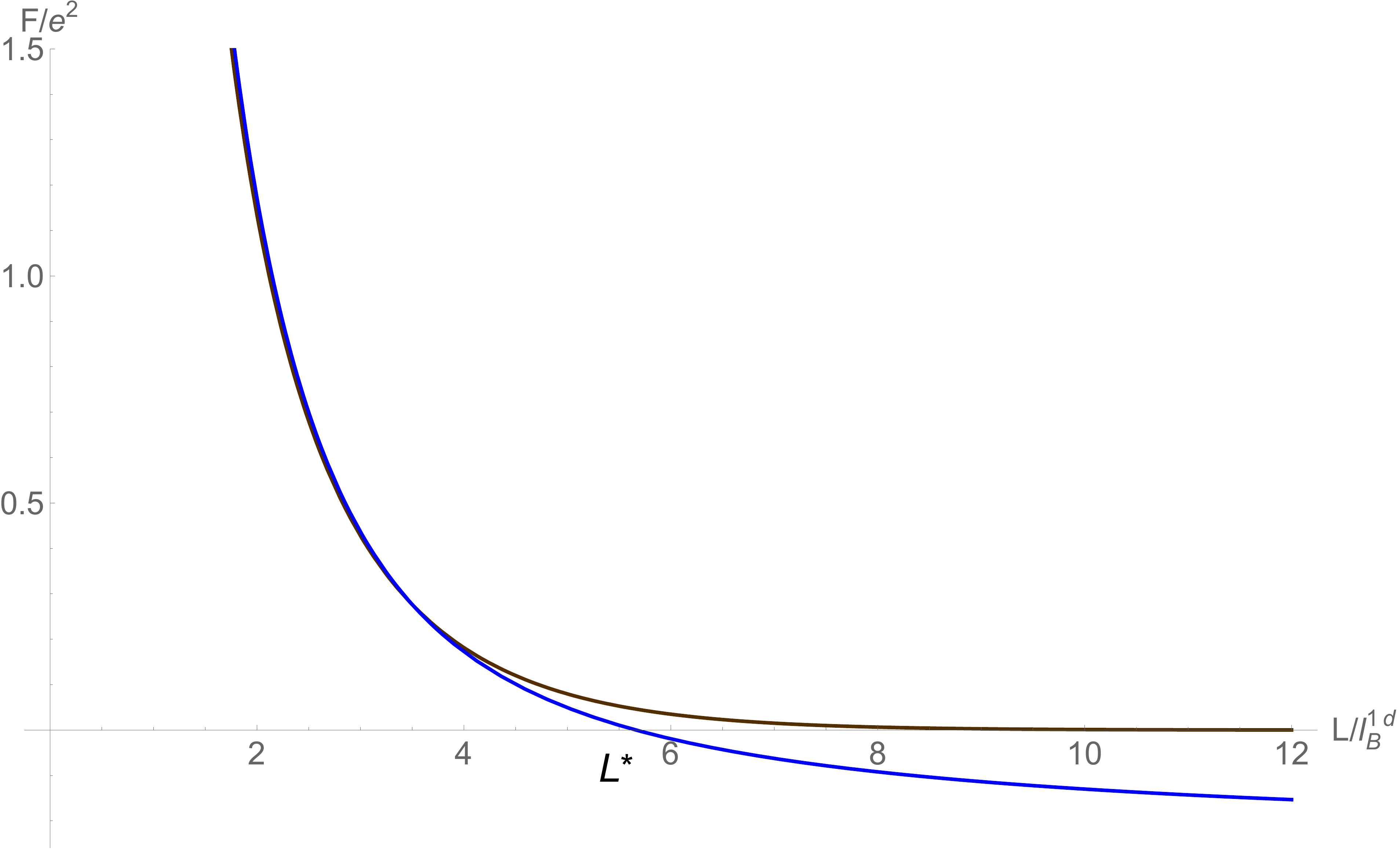}
    \end{center}
    \caption{
        \label{fig:FvsL}
        The effective force between the charges $Q$ for the cases when the
        number of counterions is even $N=10$ and odd $N=11$. This latter case
        presents the like-charge attraction phenomenon.
        }  
\end{figure}

Beyond those exact results, it is instructive to recall here a simple
argument~\cite{TellezTrizac15} which explains the like-charge attraction.
Consider first the even case $N=2p$ and that the charges $Q$ are separated by a
distance $L\gg l_B^{1d}$. The system will decouple into two parts. The charge
$Q$ located at $x=0$ will be screened by a layer of $p=N/2$ counterions and the
charge $Q$ located at $x=L$ will have its screening layer composed of the rest
of the $p$ counterions. These two entities are neutral and they weakly interact
with a repulsive force. In that situation, no like-charge attraction is
observed. In the odd case, $N=2p+1$, frustration appears in this scenario. The
counterions will try to screen each charge $Q$, but since $N$ is not divisible
by two, a misfit ion remains between two layers of $p$ counterions around each
charge $Q$. These two entities are not neutral now, they have a charge $Q+e p =
-e/2$. This situation is illustrated in figure~\ref{fig:attract1d}. The misfit
counterion between the two feels a zero electric field because the electric
field created by each entity cancels each other. This misfit counterion roams
between the two charged entities. When it is close, say to the left entity, the
charge of the entity plus the misfit ion is $+e/2$ which will create an
effective attractive force to the right entity that has charge $-e/2$. 
The effective attractive force is 
\begin{equation}
    \label{eq:force-attract}
    F \to -(e/2)^2\, \qquad \text{when } L\gg l_B^{1d}
    \,.
\end{equation}
Equation~(\ref{eq:force-attract}) gives the leading term of the force for large
distances $L$. The next-to-leading term can also be obtained by a simple
argument. The misfit ion has an available distance to move between the two
screening layers equal to $L-2l_S$ where $l_S$ is the screening layer size
around one of the charges $Q$. This layer size can be evaluated as the average
position of the $p$-th counterion of the left side. With the Laplace transform
technique, this has been computed exactly~\cite{TellezTrizac15}
\begin{equation}
    l_S = l_B^{1d}\frac{N-1}{N+1}.
\end{equation} 
The misfit ion is not subjected to any force, therefore its contribution to the
effective force can be computed as the one of an ideal gas with one particle in
a ``volume" (distance in 1d) $L-2 l_S$. It is $k_B T /(L-2l_S)$. So, the
effective force at large $L$ behaves as 
\begin{equation}
    \label{eq:force-attract-sublead}
    F \to -(e/2)^2\ + \frac{k_B T}{L-2 l_S} + o(1/L) , \qquad \text{when } L\gg l_B^{1d}
    \,.
\end{equation}
From this expression, one can evaluate the distance $L^*$ at which the force
changes from repulsive ($L<L^*$) to attractive ($L>L*$)
\begin{equation}
    L^* \simeq 4l_B^{1d}+2l_S =  \left( 4 + 2\frac{N-1}{N+1}\right) l_B^{1d}\,.
\end{equation}
As an example, figure~\ref{fig:FvsL} shows a plot of the force as a function of
the distance $L$ for the cases $N=10$ and $N=11$. For $N=10$ the force is always
repulsive, while in the odd case $N=11$ the force is repulsive at short
distance, then attractive for large distances $L$ beyond $L^* \simeq 5.67 l_B^{1d}$.

Several extensions of this 1d model have been considered, including the case
when the counterions can be also in the regions $x<0$ and $x>L$~\cite{VarelaTT17}
and when there are dielectric inhomogeneities in those different
regions~\cite{VarelaTT21}. When the dielectric constant outside (regions $x<0$
and $x>L$) is larger than inside (region $0<x<L$) it is possible to have like
charge attraction even when $N$ is even due to the effect of image charges.

In the odd case $N=2p+1$, the misfit ion plays a central role in the like-charge
attraction phenomenon. In~\cite{VarelaATT21} the dynamics of the relaxation to
a thermal equilibrium of the screening layers have been studied. Imposing an
overdamped Langevin dynamics to the counterions, it is observed that the misfit
ion rules the relaxation time $\tau$ of the system. For large distances $L\gg
l_B^{1d}$ and an odd number of particles, $\tau \propto L^2/D$, where $D$ is the
diffusion coefficient. On the other hand, for an even number of counterions, $L$
does not play any role. Instead, the relaxation time is $\tau=4 (l_B^{1d})^2
/D$. Notice also that the temperature dependency of $\tau$ changes drastically.
Recalling that $D\propto T$, we notice that in the odd case, $\tau\propto 1/T$,
whereas for $N$ even, $\tau\propto T$. Thus, when $N$ is odd, the higher the
temperature, the faster the system equilibrates. This is because the misfit ion
can roam faster between both layers. On the other hand, if $N$ is even, the
relaxation time is reduced if the temperature is low. In this case, the physical
mechanism ruling the relaxation is the formation of the layers, which will be
faster if the temperature is lower and the electrostatic coupling is larger.


\section{Two-dimensional models}
\label{sec:2d}

\subsection{Connection with integrable quantum field theories}

We move on in this section to study two-dimensional systems. Let us consider a
plasma living in a 2d plane composed of two types of particles of charges
$q_{+}>0$ and $-q_{-}<0$. The potential energy between two charges $q$ and $q'$
separated by a distance $r$ on the 2d plane is $-qq'\ln r$. In 2d, the
dimensions of a charge $q$ is $\text{energy}^{1/2}$. Comparing the electrostatic
energy to the thermal energy, we obtain an important adimensional coupling
constant of the system: $\beta q_{+} q_{-}$. We consider here a system of point
particles. Because of the attractive interaction between positive and negative
charges, the system of point particles is stable against the collapse of
oppositely charged particles only if $\beta q_{+}q_{-}<2$. We will restrict the
following analysis to that stability regime. The potential energy of the system
composed by $N_+$ positive charge located at positions $\{\r_j^{+}\}$ and $N_-$
negative charges located at $\{\r_k^{+}\}$ is 
\begin{align}
    U_{N_+,N_-} &= - q_{+}^2 \sum_{1\leq k<j\leq N_+} \ln |\r_j^{+}-\r_k^+|
    - q_{-}^2 \sum_{1\leq k<j\leq N_+} \ln |\r_j^{-}-\r_k^-|
    + q_{+}q_{-} \sum_{j=1}^{N_+} \sum_{k=1}^{N_-} \ln |\r_k^{-}-\r_j^{+}|
    \,.
\end{align}

The grand canonical partition function of the system with fugacities $z_{+}$ and
$z_{-}$ for the positive and negative particles respectively is 
\begin{equation}
    \label{eq:Xi-init}
    \Xi = \sum_{N_{+}=0}^{\infty} \sum_{N_{-}=0}^{\infty}
    \frac{z_{+}^{N_{+}}}{N_{+}!} \frac{z_{-}^{N_{-}}}{N_{-}!}
    e^{-\beta U_{N_+,N_-}} \prod_{j=1}^{N_{+}} d\r_{j}^{+} \prod_{k=1}^{N_{-}} d\r_{k}^{-}
    \,.
\end{equation} 
By using the Hubbard-Stratonovich transformation~\cite{Strat57,Hub59}, the grand
partition function~(\ref{eq:Xi-init}) can be transformed into the generating
functional of a quantum field theory 
\begin{equation}
    \Xi = \frac{Z(z_+, z_-)}{Z(0,0)},
\end{equation}
where 
\begin{equation}
    \label{eq:qft-func}
    Z(z_+, z_-) = \int \mathcal{D}\phi \, e^{-S(z_+,z_-)},
\end{equation}
with 
\begin{equation}
    \label{eq:action}
    S(z_+, z_-) = -\int \left[ \frac{\beta}{4\pi} \phi(\r)\Delta \phi(\r) 
    + z_+ e^{i\beta q_{+}\phi(\r)} 
    + z_- e^{-i\beta q_{-}\phi(\r)} 
    \right]\, d\r
    \,.
\end{equation}
Two special cases are worth mentioning. When $q_+=q_-$, the equivalent quantum
field theory is the sine-Gordon model. For $q_+=2q_-$ or $q_-=2q_+$, the quantum
field theory is the Dodd-Bullough model. In two dimensions, these two field
theories are integrable and many of their properties (spectrum, S-matrix,
expectations of exponential field, form factors, ...) have been computed
exactly~\cite{ZZ79,DV91, Z95, LZ97, L97,Smirnov1992, FLZZ98, BS01}. Using these
results, the exact equilibrium thermodynamic properties of the 2d plasma have
been obtained for the charge-symmetric situation~\cite{Samaj2000} $q_+=q_-$
and the charge-asymmetric situation~\cite{Samaj2003} with $q_+=2q_-$ or
$q_-=2q_+$. Beyond those two particular cases, the equivalent field theory has
not been solved exactly. 

Consider now two external charges $Q_1$ and $Q_2$ immersed in the plasma at $0$
and $\r$. For the following analysis to remain valid it is necessary to be in the
stability regime, not only for the plasma charges ($\beta q_{+} q_{-}<2$) but
also for the guest charges against the collapse with a plasma charge of opposite
sign 
\begin{equation}
    \label{eq:stabilityQ}
    -2 / q_{+} < \beta Q_{k} < 2/q_{-},\qquad \text{for }k\in\{1,2\}.
\end{equation}
Averaging over the thermal fluctuations of the plasma ions, the
effective interaction $G_{Q_1 Q_2}(r)$ between these two charges is defined by 
\begin{equation}
    \label{eq:Gdef}
    e^{-\beta G_{Q_1 Q_2}(r)} = \frac{\Xi[Q_1,0;Q_2,\r]/\Xi}{(\Xi[Q_1]/\Xi)\,(\Xi[Q_2]/\Xi)},
\end{equation}
where $\Xi[Q_1,0;Q_2,\r]$ is the partition function of the plasma in the
presence of the charges $Q_1$ and $Q_2$ separated by a distance $r$, while
$\Xi[Q_1]$ and $\Xi[Q_2]$ are the partition functions of the plasma with
only one guest charge $Q_1$ or $Q_2$ immersed in it. In the formalism of the
equivalent quantum field theory, the effective interaction $G_{Q_1 Q_2}(r)$ can
be cast as 
\begin{equation}
    e^{-\beta G_{Q_1 Q_2}(r)} = \frac{\left< e^{i\beta Q_1\phi(0)} e^{i\beta Q_2 \phi(\r)} \right>}{
        \left< e^{i\beta Q_1 \phi(0)} \right>
        \left< e^{i\beta Q_2 \phi(\r)} \right>
    },
\end{equation}
where the average $\left<\cdots\right>$ is taken over the fluctuations of the
field $\phi$ with the weight defined in~(\ref{eq:qft-func})
and~(\ref{eq:action}). 

For the particular cases of the charge-symmetric plasma and charge-asymmetric
plasma with $q_+/q_+$ equal to $2$ or $1/2$, using the equivalent sine-Gordon or
Dodd-Bullough theory, it is possible to obtain the asymptotic behavior of
$G_{Q_1 Q_2}(r)$ for $r\to 0$ or $r\to\infty$. For $r\to 0$ the technique is
based on the operator product expansion~\cite{Tellez2005, VarelaT21}, while for
large distances an expansion in form factors is
used~\cite{SJ02,S05anomalous,S06core,Tellez2006EPL}. Interestingly, in the
charge-asymmetric plasma with $q_+/q_-=2$ or $1/2$, the phenomenon of
like-charge attraction appears at short distance~\cite{VarelaT21} and a related
phenomenon of charge inversion is noticed at large
distances~\cite{Tellez2006EPL}. In the following, we generalize the findings of
like-charge attraction at short distance~\cite{VarelaT21} for arbitrary charge
asymmetry ($q_+/q_-$ can take arbitrary values).

\subsection{Dominant term in the effective interaction}

Starting from the definition~(\ref{eq:Gdef}) of the effective interaction $G_{Q_1
Q_2}$, we notice that the dependence on the distance $r$ between the charges is
only on the term 
\begin{equation}
    \label{eq:XiQ1Q2}
    \Xi[Q_1, 0; Q_2, \r]=\sum_{N_+=0}^{\infty} \sum_{N_+=0}^{\infty} 
    \frac{z_+^{N_+}}{N_+!}\frac{z_-^{N_-}}{N_-!}
    Z_{ N_+, N_-} [Q_1, 0; Q_2, \r].
\end{equation}
In~(\ref{eq:XiQ1Q2}) appears a sum of configurational integrals of the two fixed
charges $Q_1$ and $Q_2$ in the presence of $N_+$ positive charges and $N_-$
positive charges 
\begin{equation}
    Z_{N_+, N_-} [Q_1, 0; Q_2, \r] = \int 
    e^{-\beta {\cal U}_{N_+, N_-}[Q_1, 0; Q_2, \r]}
    \,
    \prod_{j=1}^{N^+} d\r_j^+
    \prod_{k=1}^{N^-} d\r_k^-
    ,
\end{equation}
with 
\begin{align}
    {\cal U}_{N_+, N_-}[Q_1, 0; Q_2, \r]
    = & -Q_1 Q_2 \ln |\r|  \\
    &- Q_1 q_{+} \sum_{j=1}^{N_+} \ln |\r_{j}^+|
    + Q_1 q_{-} \sum_{k=1}^{N_-} \ln |\r_{k}^-| \\
    & - Q_2 q_{+} \sum_{j=1}^{N_+} \ln |\r-\r_{j}^+|
    + Q_2 q_{-} \sum_{k=1}^{N_-} \ln |\r-\r_{k}^-| \\
    & + U_{N_+,N_-}
    \,.
\end{align}
The explicit dependence on $r=|\r|$ of $Z_{ N_+, N_-} [Q_1, 0; Q_2, \r]$ can be
obtained by simple dimensional analysis. Let $(r,\theta)$ be the polar
coordinates of $\r$. Using complex coordinates, one can make a change of
variables on the plasma charges positions to rescale them by $r$ and rotate them
so that the charge $Q_2$ is formally located at $1$. Explicitly, $z_{j}^{+}=
(r_{j}^{+}/r) e^{i(\theta_{j}^{+}-\theta)}$ and $z_{k}^{-}= (r_{k}^{-}/r)
e^{i(\theta_{k}^{-}-\theta)}$. Then 
\begin{equation}
    Z_{N_+, N_-} [Q_1, 0; Q_2, \r] = r^{\beta Q_1 Q_2 + f(Q_1+Q_2; N_+,N_-)}
    Z_{N_+, N_-} [Q_1, 0; Q_2, 1],
\end{equation}  
with 
\begin{align}
    f(Q, N_+,N_-) = 
    N_{+}(2-\beta q_{+}^2/2)
    +
    N_{-}(2-\beta q_{-}^2/2)
    +
    \beta Q(q_{+} N_{+}-q_{-}N_{-})
    + \beta (q_{+} N_{+}-q_{-}N_{-})^2/2
    ,
\end{align}
and 
\begin{align}
    Z_{N_+, N_-} [Q_1, 0; Q_2, 1] =
    \int &
    \frac{\prod_{j=1}^{N_+} |z_j^{+}|^{\beta q_ {+}Q_1} |1-z_j^{+}|^{\beta q_{+}Q_2}}{
        \prod_{k=1}^{N_-} |z_k^{-}|^{\beta q_ {-}Q_1} |1-z_k^{-}|^{\beta q_{-}Q_2}
        }
        \nonumber \\
    & \times \frac{
        \prod_{1\leq j < k \leq N_{+}}|z_{j}^{+}-z_{k}^+|^{\beta q_{+}^2}
        \prod_{1\leq j < k \leq N_{-}}|z_{j}^{-}-z_{k}^-|^{\beta q_{-}^2}
    }{
        \prod_{j=1}^{N_+} \prod_{k=1}^{N_-} |z_{j}^{+} - z_{k}^{-}|^{\beta q_{+}q_{-}}
    }
    \prod_{j=1}^{N_+}d^2 z_{j}^{+}
    \prod_{k=1}^{N_-}d^2 z_{k}^{-}
    \,.
\end{align}

From the previous expressions, we notice that $\exp(-\beta G_{Q_1 Q_2}(r))$ is
given by a sum of terms with power law dependency on the distance between the
charges: $r^{\beta Q_1 Q_2 + f(Q_1+Q_2,N_+,N_-)}$. Physically this is
interpreted as the contributions to the effective potential when the charges
$Q_1$ and $Q_2$ are successively screened by $N_{+}$ positive charges and $N_-$
negative charges of the plasma. In the context of the equivalent field theory,
this is the product operator expansion used in~\cite{Tellez2005,VarelaT21}. For
small distances $r\to 0$, one can analyze the behavior of the effective
potential by determining the dominant term in this expansion as follows.

For sufficiently small values of $\beta$, $|Q_1|$ and $|Q_2|$ we have
$f(Q_1,Q_2,N_+,N_-)\geq f(Q_1,Q_2,0,0)=0$. Therefore the dominant term
in~(\ref{eq:XiQ1Q2}) for $r\to 0$ is obtained when $N_+=0$ and $N_-=0$,
\begin{equation}
    \Xi[Q_1, 0; Q_2, \r] \sim e^{\beta Q_1 Q_2},
\end{equation}
and the effective interaction behaves like a bare Coulomb potential 
\begin{equation}
    G_{Q_1 Q_2}(r) \sim -Q_1 Q_2 \ln r.
\end{equation}
However as $\beta$, $|Q_1|$ and $|Q_2|$ increase, it is possible that
$f(Q_1,Q_2,N_+,N_-)<0$ for some values of $N_+$ and $N_-$. This will induce a
change in the short distance behavior of $G_{Q_1 Q_2}(r)$ from the bare Coulomb
potential to a different behavior. These changes were first predicted
in\cite{HansenViot85} for the correlation functions of a homogenous plasma by
an analysis in the canonical ensemble. Here we focus on the situation with
the two guest charges $Q_1$ and $Q_2$ and derive those changes working in the
grand canonical ensemble setting which simplifies the analysis.  

To fix the ideas, consider $Q_1<0$ and $Q_2<0$. Let us find when a term $f(Q,
N_{+}+1, N_{-})$ dominates over a term $f(Q, N_{+}, N_{-})$ where $Q=Q_1+Q_2$.
The dominant term will be the smaller one. Since  
\begin{equation}
    f(Q, N_{+}, N_{-})-f(Q, N_{+}+1, N_{-}) = 2 + \beta q_{+} (Q+q_{+}N_{+} - q_{-} N_{-})
    \,.
\end{equation}
Then $f(Q, N_{+}+1, N_{-})$ dominates over $f(Q, N_{+}, N_{-})$ when 
\begin{equation}
    \beta q_{+} Q < -2 - \beta q_{+} (q_+ N_+ - q_- N_-).
\end{equation}
Notice that this corresponds exactly to the stability threshold for a negative
cluster of charge $Q+q_+ N_+ -q_- N_-$:
\begin{equation}
    \label{eq:threshold-changeNN}
    \beta (Q + q_+ N_+ - q_- N_-) < -2/q_{+}.
\end{equation}
However, since the charges $Q_1$ and $Q_2$ are not located in the same position
$r\neq 0$ it is possible to be in a situation
where~(\ref{eq:threshold-changeNN}) occurs but the individual charges are still
in their stability regime~(\ref{eq:stabilityQ}) guaranteeing that
$\Xi[Q_1,0;Q_2,\r]$ does not diverges. The physical interpretation is that a
term with $r$-power law $\beta Q_1 Q_2 + f(Q, N_{+}, N_{-})$ dominates as long
as the cluster of negative charge $Q + q_+ N_+ - q_- N_-$ is bounded below by
$-2/q_+$. As soon as it crosses this value, an additional ion $q_+$ comes into the
picture to strengthen the cluster screening and the $r$-power law will change to
$\beta Q_1 Q_2 + f(Q, N_{+}+1, N_{-})$. To complete the analysis we have to
compare also with the term with $N_{+}-1$ positive charges and $N_{-}$ negative
charges. This follows simply by changing $N_+$ by $N_+-1$ in the previous
analysis, to conclude that the contribution from the cluster with charge $Q +
q_+ N_+ -q_- N_-$ is dominant over other clusters with different values of $N_+$
whenever 
\begin{equation}
    \label{eq:Qneg_dom}
    -2 - \beta q_{+}(q_+ N_+ - q_- N_-) < 
    \beta q_{+} Q < 
    -2 - \beta q_{+} (q_+ (N_+ - 1) - q_- N_-) < 0.
\end{equation}
When $\beta q_{+} Q$ becomes smaller than the lower bound, the term with an
additional $q_+$ ion becomes dominant (cluster of charge $Q + q_+ (N_+ + 1) -q_-
N_-$). On the other hand when $\beta q_{+} Q$ was larger than the upper bound,
the dominant term was for a cluster  of charge $Q + q_+ (N_+ - 1) -q_-
N_-$.

Similarly, if we consider positive guest charges $Q_1>0$ and $Q_2>0$, the
regime where a cluster of charge $Q+ q_- N_- -q_+ N_+$ is dominant is when 
\begin{equation}
    \label{eq:Qpos_dom}
    0< 2 + \beta q_- (q_- (N_- - 1) - q_+ N_+) < 
    \beta q_{-} Q < 
    2 + \beta q_- (q_- N_- - q_+ N_+).
\end{equation}
When $\beta q_{-} Q$ was below the lower bound, the dominant term was given by a
cluster of charge $Q - q_- (N_- - 1) + q_+ N_+$, and when it surpasses the upper
bound, the dominant term is given by a cluster with an additional negative ion
(charge $Q-q_- (N_- + 1) + q_+ N_+$).

\begin{figure}
    \begin{center}
        \includegraphics[width=0.9\textwidth]{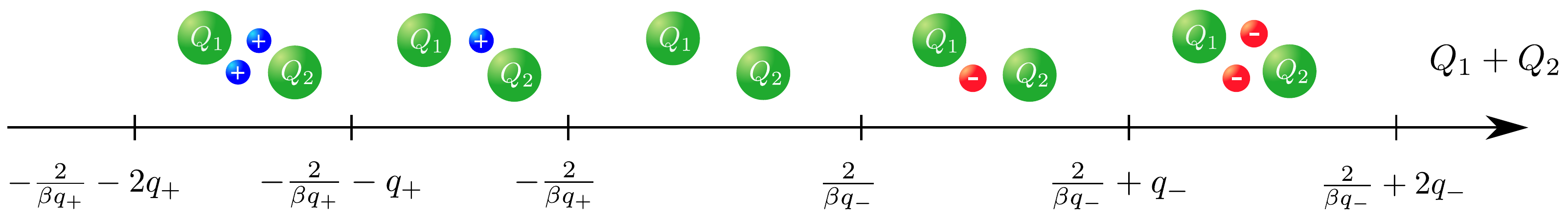}
    \end{center}
    \caption{
        \label{fig:dominant-term-order}
        Graphical illustration of the dominant term in the effective interaction
        depending on the sum $Q_1+Q_2$ of the guest charges. }  
\end{figure}

Continuing with the case of positive guest charges, let us suppose that the
condition~(\ref{eq:Qpos_dom}) is satisfied so the contribution from a cluster of
charge $Q-q_- N_- + q_+ N_+$ dominates over other clusters with a different
number of negative ions $N_-' \neq N_-$. How does the contribution of this
cluster compare with the one from another cluster with a different number $N_+'$
of positive ions? Let us compare it with a cluster with $N_+ - 1$ positive ions.
We have 
\begin{equation}
    f(Q, N_+, N_-) - f(Q, N_+-1, N_-) = 2 + \beta q_{+} (Q + q_{+}(N_{+}-1)-q_{-}N_-) + 2
    .
\end{equation}
Since $Q$ satisfies~(\ref{eq:Qpos_dom}), we have 
\begin{equation}
    f(Q, N_+, N_-) - f(Q, N_+-1, N_-) > (2-\beta q_+ q_-)q_+/q_- +  1 > 1 >0
    ,
\end{equation}
where we used the stability condition $\beta q_+ q_- <2$. Therefore the
contribution from a cluster with $N_{+}-1$ positive will dominate. Repeating
this recursively we deduce that the cluster that will dominate is the one without
positive ions ($N_+=0$) and $N_{-}$ negative ions. In conclusion, to obtain the
dominant behavior we need to consider only clusters with ions of opposite charge to
the guest charges. For $Q_1>0$ and $Q_2>0$, this means that the successive
changes in the dominant behavior of the effective interaction will be given by
the contributions of clusters of charge $Q_1+Q_2 -q_{-} N_{-}$ when 
\begin{equation}
\label{eq:Qpos_domN0}
    0< 2 + \beta q_-^2 (N_- - 1)  < 
    \beta q_{-} ( Q_1+Q_2 )< 
    2 + \beta q_-^2 N_- . 
\end{equation}
For negative guest charges, it is the cluster of charge $Q_1+Q_2 +q_{+} N_{+}$
that dominates when 
\begin{equation}
    \label{eq:Qneg_domN0}
    -2 - \beta q_+^2 N_+  < 
    \beta q_{+} (Q_1 + Q_2) < 
    -2 - \beta q_+^2 (N_+ - 1)  < 0.
\end{equation}
The effective interaction between $Q_1$ and $Q_2$ is given by 
\begin{equation}
    \label{eq:domG}
    G_{Q_1 Q_2}(r) \sim (Q_1 Q_2 +\beta^{-1}f(Q_1+Q_2, N_{+}, N_{-})) \ln r
    ,
\end{equation}
with $N_{+}= 0$ and a given value of $N_-\geq 0$ when $Q_1+Q_2$
satisfies~(\ref{eq:Qpos_domN0}), or $N_{-}=0$ and a given value of $N_{+}\geq 0$
when~(\ref{eq:Qneg_domN0}) is satisfied. This is illustrated in
figure~\ref{fig:dominant-term-order}.
%


\subsection{Like-charge attraction}

To analyze the possibility of like-charge attraction, one has to determine
the sign of $\beta Q_1 Q_2 +f(Q_1+Q_2, N_{+}, N_{-})$ in the dominant
behavior~(\ref{eq:domG}) of $G_{Q_1 Q_2}(r)$ when $Q_1$ and $Q_2$ have the same
sign. Without loss of generality let us consider the case $Q_1>0$ and $Q_2>0$.
The opposite case follows by interchanging the roles of $q_{+}$ and $q_{-}$. The
bare interaction $-Q_1 Q_2 \ln r$ is repulsive. Let us explore the regions after
the first change of behavior of $G_{Q_1 Q_2}$, when $N_{-}\geq 1$ negative ions
and $N_{+}=0$ positive ions form a cluster with $Q_1$ and $Q_2$. Let us define 
\begin{equation}
    \tilde{Q}_k = \beta Q_k q_{-} /2,\quad \text{for } k\in\{1,2\},
\end{equation}
and 
\begin{equation}
    \tilde{q}_{-} = \beta q_{-}^2 /2 .
\end{equation}
Because of the stability condition $(\tilde{Q}_1, \tilde{Q}_2)\in[0,1)^2$ and
$\tilde{q}_{-}\in[0, q_{-}/q_{+})$. With this notation, the
condition~(\ref{eq:Qpos_domN0}) can be written as 
\begin{equation}
    \label{eq:bands}
    1 + (N_- -1)\tilde{q}_{-} < \tilde{Q}_1 + \tilde{Q}_2 < 1 + N_{-}\tilde{q}_{-}.
\end{equation}
When this is satisfied, the effective interaction at short distances is given by
\begin{equation}
    \label{eq:GeffNp0Nm}
    \beta G_{Q_1 Q_2}(r)\sim (\beta Q_1 Q_2 +f(Q_1,Q_2,0,N_{-})) \ln r
    .
\end{equation}
This interaction is repulsive as long as $\beta Q_1 Q_2 + f(Q_1,Q_2,0,N_{-})>0$
and it will become attractive as soon as that quantity becomes negative. This
condition can be cast as 
\begin{equation}
    \label{eq:LCA}
    g(\tilde{Q}_1, \tilde{Q}_2) < g(0,1+\tilde{q}_{-}(N_{-}-1)/2),
\end{equation}
where we defined 
\begin{equation}
    g(\tilde{Q}_1,\tilde{Q}_2)=
    (\tilde{Q}_1-N_{-}\tilde{q}_{-})(\tilde{Q}_2-N_{-}\tilde{q}_{-})
    \,.
\end{equation}
The separatrix between the repulsive and attractive regions is defined by the
hyperbolas
\begin{equation}
    \label{eq:hyperbolas}
    g(\tilde{Q}_1,\tilde{Q}_2)=g(0,1+\tilde{q}_{-}(N_{-}-1)/2),
\end{equation}
which have asymptotes at $\tilde{Q}_1=N_{-}\tilde{q}_{-}$ and
$\tilde{Q}_2=N_{-}\tilde{q}_{-}$ and they pass thru the points $(\tilde{Q}_1,
\tilde{Q}_2)=(0,1+\tilde{q}_{-}(N_{-}-1)/2)$ and $(\tilde{Q}_1,
\tilde{Q}_2)=(1+\tilde{q}_{-}(N_{-}-1)/2,0)$. To determine if like-charge
attraction is possible, we need to find if the region defined
by~(\ref{eq:bands}) and~(\ref{eq:LCA}) has a non empty intersection with the stability region
$[0,1)^2$.

\begin{figure}
    \begin{center}
        \includegraphics[width=0.7\textwidth]{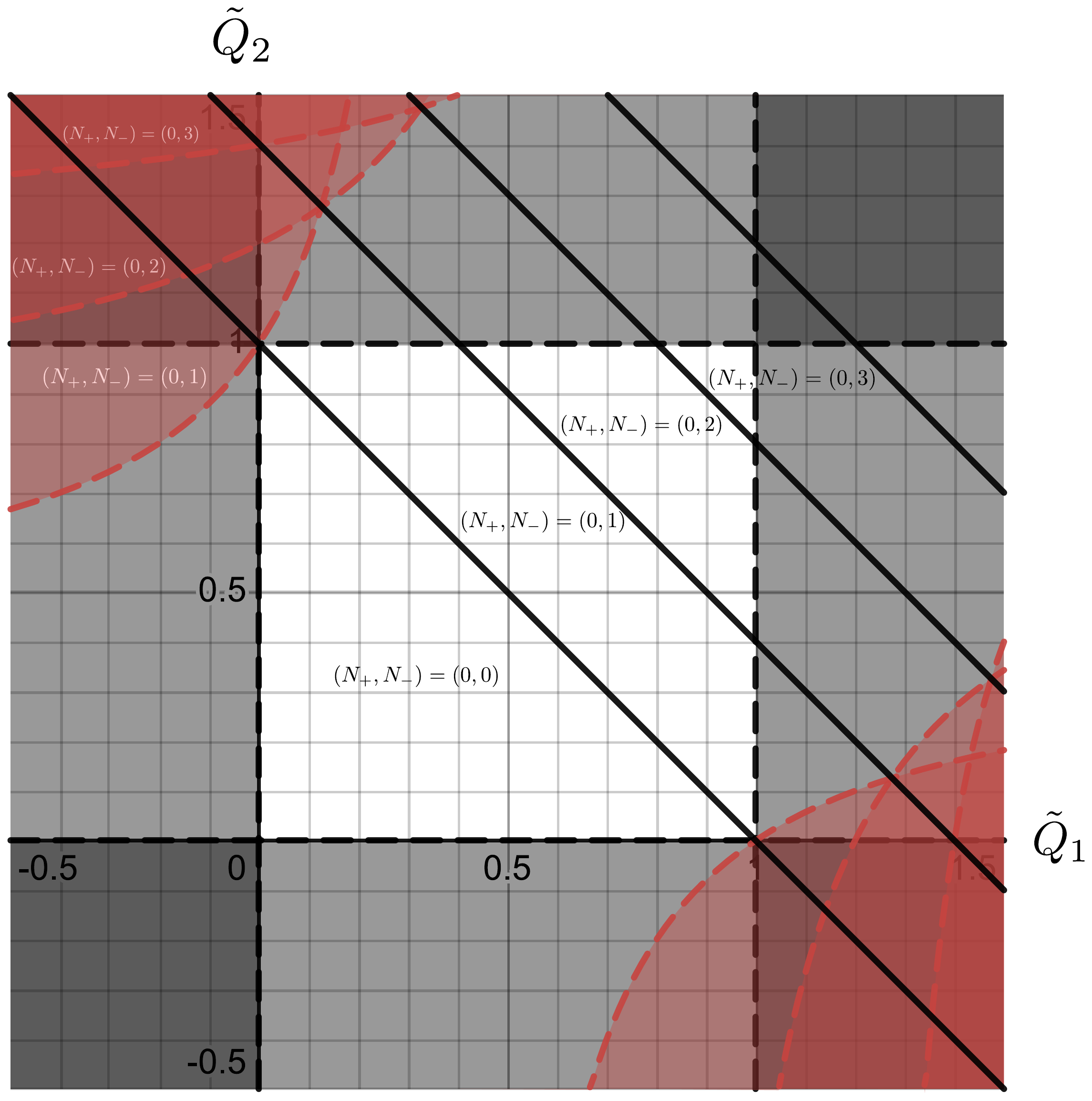}
    \end{center}
    \caption{
        \label{fig:non_lca_q0_4}
        When $\tilde{q}_{-}<1$ (here $\tilde{q}_{-}=0.4$) the stability region
        in white does not intersect the region of possible like-charge
        attraction in red defined by~(\ref{eq:LCA}). Since greyed zone is
        forbidden by the stability condition~(\ref{eq:stabilityQ}), the
        effective interaction remains always repulsive (white region). Notice
        that there are several changes in the effective interaction behavior
        depending on the bands that are shown in the upper triangle, each
        corresponding to a cluster of the guest charges with $N_{-}$ negative
        ions.
        }  
\end{figure}

For this we need to distinguish two cases if $\tilde{q}_{-}<1$ or
$\tilde{q}_{-}>1$. The situation when $\tilde{q}_{-}<1$ is shown in
figure~\ref{fig:non_lca_q0_4}. The stability domain $[0,1)^2$ is divided into
several regions depending on the behavior of the effective potential. Below the
diagonal $\tilde{Q}_1+\tilde{Q}_2 = 1$, the effective interaction behaves like
the bare Coulomb potential. This is the lower triangle of the square domain
$[0,1)^2$. The upper triangle is divided by diagonal bands defined
in~(\ref{eq:bands}). In each of these bands, the effective interaction behaves
as~(\ref{eq:GeffNp0Nm}). For $\tilde{q}_{-}<1$, several bands occupy the upper
triangle up to $N_{-} < 1+ 1/\tilde{q}_{-}$. For each of these zones, there is
pair of hyperbolas defined by~(\ref{eq:hyperbolas}) for the corresponding value
of $N_{-}$. If $\tilde{q}_{-}<1/2$, the l.h.s of~\ref{eq:hyperbolas} is negative
$g(0,1+\tilde{q}_{-}(N_{-}-1)/2)<0$, and the corresponding hyperbolas lie on the
left upper and right lower quadrants. Therefore the zone defined
by~(\ref{eq:LCA}) never intersects the stability region, see
figure~\ref{fig:non_lca_q0_4}. 

\begin{figure}
    \begin{center}
        \includegraphics[width=0.7\textwidth]{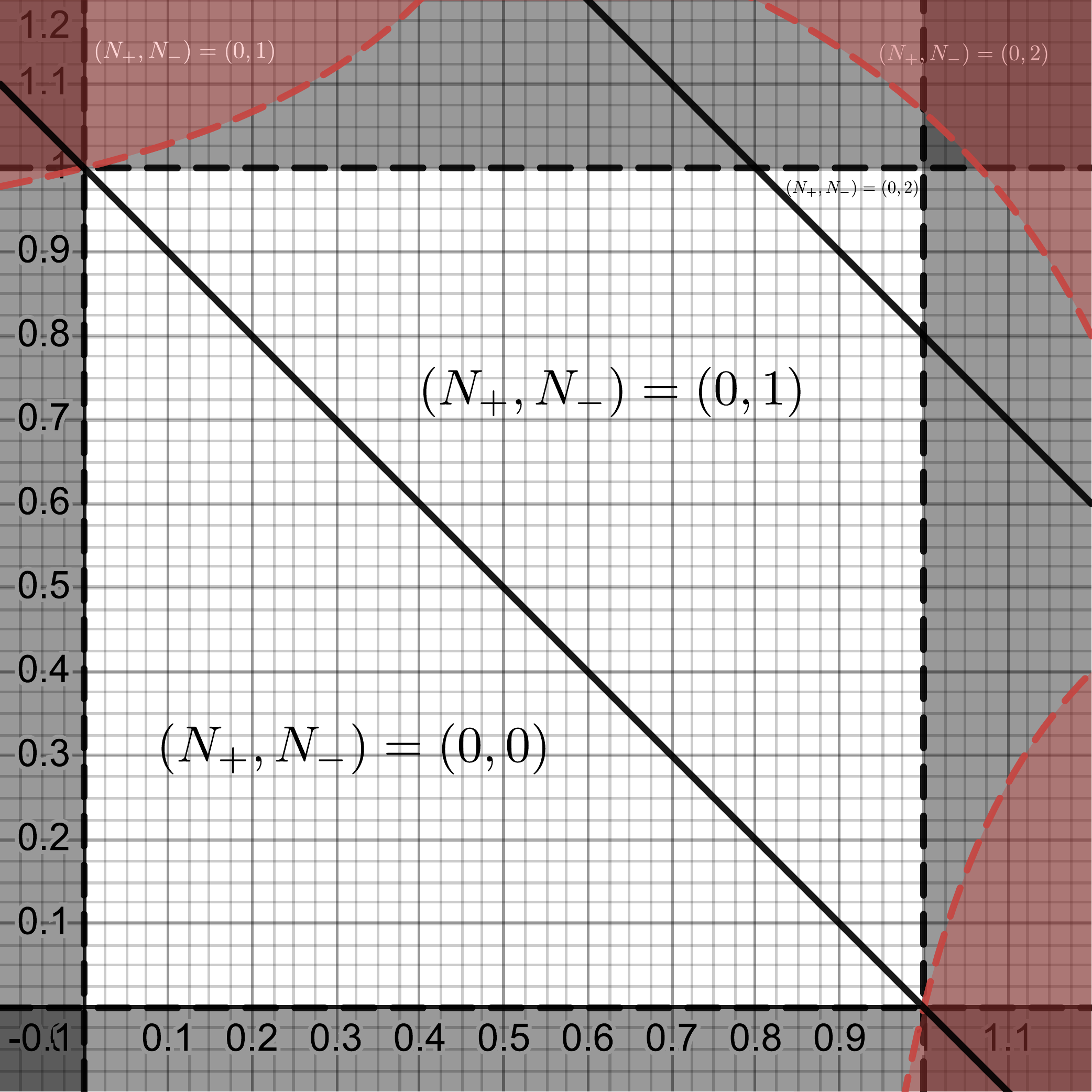}
    \end{center}
    \caption{
        \label{fig:non_lca_q0_8}
        When $1/2<\tilde{q}_{-}<1$ (here $\tilde{q}_{-}=0.8$) one pair of
        hyperbolas~(\ref{eq:hyperbolas}) have switch sides. Nevertheless, the
        stability region in white does not intersect the region of possible
        like-charge attraction in red defined by~(\ref{eq:LCA}). Since the greyed
        zone is forbidden by the stability condition~(\ref{eq:stabilityQ}), the
        effective interaction remains always repulsive (white region). }  
\end{figure}

When $1/2< \tilde{q}_{-}< 1$, it is possible that some hyperbolas switch
sides to the upper right corner and the lower left corner. This happens when the
l.h.s of~\ref{eq:hyperbolas} is positive $g(0,1+\tilde{q}_{-}(N_{-}-1)/2)>0$.
This corresponds to $N_{-}> (2/\tilde{q}_-)-1$. Nevertheless, the hyperbolas
remain outside the stability region as long as $\tilde{q}_{-}<1$ as seen in
figure~\ref{fig:non_lca_q0_8}. In conclusion, as illustrated in
figures~\ref{fig:non_lca_q0_4} and~\ref{fig:non_lca_q0_8} there is never an
intersection between the stability region and the region defined
by~(\ref{eq:LCA}). The regions where~(\ref{eq:LCA}) is satisfied lie always in
the forbidden zone outside the stability domain. In conclusion, the effective
potential remains always repulsive when $\tilde{q}_{-}<1$.

\begin{figure}
    \begin{center}
        \includegraphics[width=0.7\textwidth]{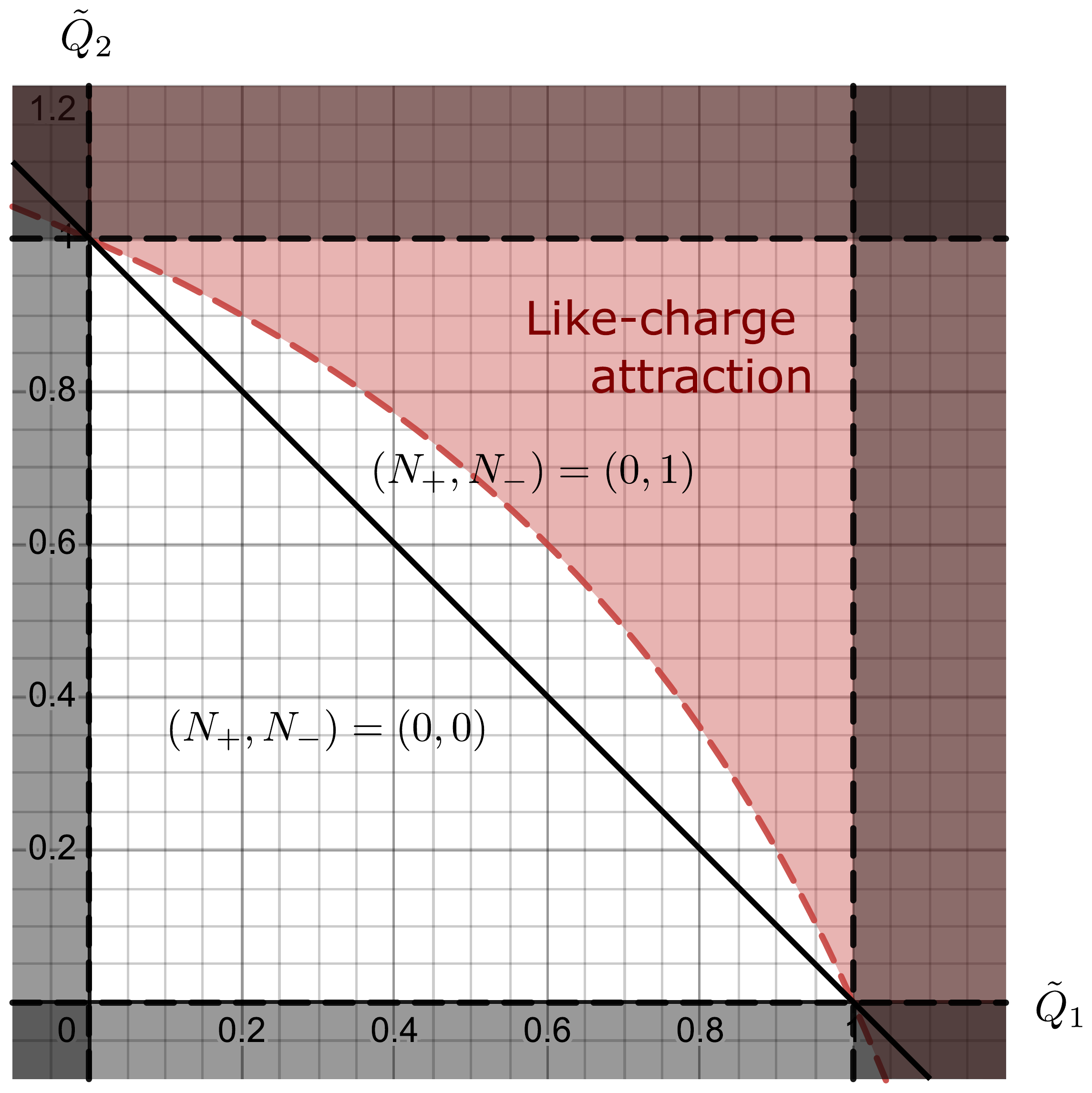}
    \end{center}
    \caption{
        \label{fig:lca}
        When $\tilde{q}_{-}>1$ (here $\tilde{q}_{-}=1.8$) the stability region
        in white does intersect the region of possible like-charge attraction in
        red defined by~(\ref{eq:LCA}). The effective interaction can change from
        a repulsive one (white region) to an attractive one (red region). 
        The greyed zone is forbidden by the stability
        condition~(\ref{eq:stabilityQ}). }  
\end{figure}

For $\tilde{q}_{-}\to 1^{-}$, the distance of the hyperbola to the point
$(\tilde{Q}_1,\tilde{Q}_2)=(1,1)$ tends to $0$, signaling the possibility of
like-charge attraction if $\tilde{q}_{-}>1$. In that situation, there is only
one band~(\ref{eq:bands}) for $N_{-}=1$ in the stability region (see
figure~\ref{fig:lca}), therefore only one change in the behavior of the
effective interaction. Now there is a non-null intersection of that upper band
and the region~(\ref{eq:LCA}) for $N_{-}=1$. So, when $\tilde{q}_{-}>1$,
like-charge attraction is possible, as shown in figure~\ref{fig:lca}. Notice
however that this situation is only possible if $q_{-}>q_{+}$ because due to the
stability condition $\tilde{q}_{-}\in[0, q_{-}/q_{+})$.

Since the hyperbola that delimits the like-charge attraction region has
asymptotes at $\tilde{Q}_1=\tilde{q}_{-}$ and $\tilde{Q}_2=\tilde{q}_{-}$, we
deduce that the larger $\tilde{q}_{-}$ is, the larger the like-charge attraction
region will be. But the maximum value for $\tilde{q}_{-}$ is $q_{-}/q_{+}$.
Therefore to increase the possibility of like-charge attraction the ratio
$q_{-}/q_{+}$ should be as large as possible. The charge-asymmetry is essential
to observe the like-charge phenomenon. In the case of a charge-symmetric plasma
of point particles, no like-charge attraction is observed.

The analysis for the case of negative guest charges is obtained from the
previous one by interchanging the roles of $q_{-}$ and $q_{+}$. For that
situation it is convenient to define $\hat{Q}_{1,2}=\beta q_{+} Q_{1,2}/2$ and
$\hat{q}_{+}=\beta q_{+}^2/2$. The possibility of like charge attraction is when  
\begin{equation}
    (\hat{Q}_1+\hat{q}_{+})(\hat{Q}_2+\hat{q}_{+})
    <
    \hat{q}_{+}(-1+\hat{q}_{+})
\end{equation}
which is only possible if $\hat{q}_{+} > 1$. This situation can only be achieved
if $q_{+}>q_{-}$ because the stability condition restricts $\hat{q}_{+}\in[0,
q_{+}/q_{-})$. Therefore, in general, it is only possible to have like-charge
attraction between one type of charges: positives if $q_{-}>q_{+}$ or the
negative ones if $q_{+}>q_{-}$, but not both at the same time. So, the
like-charge attraction phenomenon can occur when the point guest charges are
screened by the ions of larger valency of the plasma.

All the previous analysis is restricted to the case of point particles, which is
the reason to restrict the values of the charges to the stability region. To
study the possibility of like-charge attraction beyond the stability region, it
is necessary to introduce a non-electric short-distance repulsive potential
between charges. This breaks downs the scale invariance of the pair potential
and invalidates our analysis. The study of the region outside the stability
regime requires special techniques~\cite{S06core, Tellez07}, and the properties of
the system change drastically compared to the case of point particles
in the stability region. Therefore we will not speculate here on what could
happen beyond the stability region. 

\subsection{Repulsion of charges of opposite sign?}

A complementary question to the like-charge attraction dilemma is whether is it
possible for opposite charges to repel each other? The answer is no, at least in
the stability regime. This is because there are no changes in the effective
interaction from the bare Coulomb potential if the charges have opposite signs.
Indeed, suppose that $Q_1>0$ and $Q_2<0$, with $Q_1+Q_2>0$. For the effective
potential $G_{Q_1 Q_2}$ to change from the bare Coulomb interaction to a
different form, it is needed that $\tilde{Q}_1 + \tilde{Q}_2 > 1$. But this
implies that $\tilde{Q}_1 > 1 + |\tilde{Q}_2| > 1$. Therefore $\tilde{Q}_1$ will
be outside the stability region for this to occur. In conclusion, the effective
interaction between oppositely charged particles behaves always as the bare
attractive Coulomb potential at short distances $G_{Q_1 Q_2}(r)\sim -Q_1 Q_2 \ln
r$.

\section{Conclusion}

We have reviewed the mechanism of like-charge attraction in 1d and 2d exactly
solvable Coulomb systems. Although the technical details differ depending on the
dimension, the key mechanism is a three-body interaction between the two
like-charges and an oppositely charged ion of the electrolyte or plasma. The
charge asymmetry of the plasma increases the like-charge attraction effect.
These 1d and 2d models provide a strong base to understand this intriguing
phenomenon in 3d systems. In particular, like-charge attraction for parallel
plates in an electrolyte in three dimensions has been put in evidence in the
strong coupling regime and explained~\cite{SamajTrizac11} using the one-single
particle picture presented in the 1d model in section~\ref{sec:1d}. This has
important practical applications, for example in understanding the cement
cohesion~\cite{cement21}.


This work was supported by Fondo de Investigaciones, Facultad de Ciencias,
Universidad de los Andes INV-2021-128-2267. I thank Lucas Varela, Sergio
Andraus, and Emmanuel Trizac for interesting discussions.

\bibliography{biblio}

\providecommand{\url}[1]{\texttt{#1}}
\providecommand{\urlprefix}{}
\providecommand{\foreignlanguage}[2]{#2}
\providecommand{\Capitalize}[1]{\uppercase{#1}}
\providecommand{\capitalize}[1]{\expandafter\Capitalize#1}
\providecommand{\bibliographycite}[1]{\cite{#1}}
\providecommand{\bbland}{and}
\providecommand{\bblchap}{chap.}
\providecommand{\bblchapter}{chapter}
\providecommand{\bbletal}{et~al.}
\providecommand{\bbleditors}{editors}
\providecommand{\bbleds}{eds: }
\providecommand{\bbleditor}{editor}
\providecommand{\bbled}{ed.}
\providecommand{\bbledition}{edition}
\providecommand{\bbledn}{ed.}
\providecommand{\bbleidp}{page}
\providecommand{\bbleidpp}{pages}
\providecommand{\bblerratum}{erratum}
\providecommand{\bblin}{in}
\providecommand{\bblmthesis}{Master's thesis}
\providecommand{\bblno}{no.}
\providecommand{\bblnumber}{number}
\providecommand{\bblof}{of}
\providecommand{\bblpage}{page}
\providecommand{\bblpages}{pages}
\providecommand{\bblp}{p}
\providecommand{\bblphdthesis}{Ph.D. thesis}
\providecommand{\bblpp}{pp}
\providecommand{\bbltechrep}{}
\providecommand{\bbltechreport}{Technical Report}
\providecommand{\bblvolume}{volume}
\providecommand{\bblvol}{Vol.}
\providecommand{\bbljan}{January}
\providecommand{\bblfeb}{February}
\providecommand{\bblmar}{March}
\providecommand{\bblapr}{April}
\providecommand{\bblmay}{May}
\providecommand{\bbljun}{June}
\providecommand{\bbljul}{July}
\providecommand{\bblaug}{August}
\providecommand{\bblsep}{September}
\providecommand{\bbloct}{October}
\providecommand{\bblnov}{November}
\providecommand{\bbldec}{December}
\providecommand{\bblfirst}{First}
\providecommand{\bblfirsto}{1st}
\providecommand{\bblsecond}{Second}
\providecommand{\bblsecondo}{2nd}
\providecommand{\bblthird}{Third}
\providecommand{\bblthirdo}{3rd}
\providecommand{\bblfourth}{Fourth}
\providecommand{\bblfourtho}{4th}
\providecommand{\bblfifth}{Fifth}
\providecommand{\bblfiftho}{5th}
\providecommand{\bblst}{st}
\providecommand{\bblnd}{nd}
\providecommand{\bblrd}{rd}
\providecommand{\bblth}{th}
\begin{thebibliography}{10}

\bibitem{Ise83}
N.~Ise, T.~Okubo, M.~Sugimura, K.~Ito, H.~J. Nolte, {\it The Journal of
  Chemical Physics} \textbf{1983}, {\it 78}, 536--540.

\bibitem{Larsen97}
Larsen A., Grier D., {\it Nature} \textbf{1997}, {\it 385}, 230--233.

\bibitem{Jensen98}
N.~Grønbech-Jensen, K.~M. Beardmore, Ph. Pincus, {\it Physica A}
  \textbf{1998}, {\it 261}, 74--81.

\bibitem{Linse99}
P.~Linse, V.~Lobaskin, {\it Phys. Rev. Lett.} \textbf{1999}, {\it 83},
  4208--4211.

\bibitem{Neu99}
J.~C. Neu, {\it Phys. Rev. Lett.} \textbf{1999}, {\it 82}, 1072--1074.

\bibitem{TrizacRaimbault99}
E.~Trizac, J.-L. Raimbault, {\it Phys. Rev. E} \textbf{1999}, {\it 60},
  6530--6533.

\bibitem{Trizac2000}
E.~Trizac, {\it Phys. Rev. E} \textbf{2000}, {\it 62}, R1465--R1468.

\bibitem{Squires2000}
T.~M. Squires, M.~P. Brenner, {\it Phys. Rev. Lett.} \textbf{2000}, {\it 85},
  4976--4979.

\bibitem{Rouzina96}
I.~Rouzina, V.~A. Bloomfield, {\it J. Chem. Phys} \textbf{1996}, {\it 100},
  9977--9989.

\bibitem{Shklovskii02}
A.~Yu. Grosberg, T.~T. Nguyen, B.~I. Shklovskii, {\it Rev. Mod. Phys.}
  \textbf{2002}, {\it 74}, 329--345.

\bibitem{Levin02}
Y.~Levin, {\it Reports on Progress in Physics} \textbf{2002}, {\it 65}, 1577.

\bibitem{Netz01}
A.~G. Moreira, R.~R. Netz, {\it Phys. Rev. Lett.} \textbf{2001}, {\it 87},
  078301.

\bibitem{SamajTrizac11}
L.~\ifmmode~\check{S}\else \v{S}\fi{}amaj, E.~Trizac, {\it Phys. Rev. Lett.}
  \textbf{2011}, {\it 106}, 078301.

\bibitem{TrizacSamaj13}
E.~Trizac, L.~\ifmmode~\check{S}\else \v{S}\fi{}amaj, {\it Physics of Complex
  Colloids}, IOS Press Ebooks, \textbf{2013}, \bblchapter{} Like-charge
  colloidal attraction: A simple argument, \bblpp{}. 61--73.

\bibitem{Lenard61}
A.~Lenard, {\it Journal of Mathematical Physics} \textbf{1961}, {\it 2},
  682--693.

\bibitem{Lenard62}
S.~F. Edwards, A.~Lenard, {\it Journal of Mathematical Physics} \textbf{1962},
  {\it 3}, 778--792.

\bibitem{Prager62}
S.~Prager, {\it Adv. Chem. Phys.} \textbf{1962}, {\it 4}, 201.

\bibitem{Dean09}
D.~S. Dean, R.~R. Horgan, A.~Naji, R.~Podgornik, {\it The Journal of Chemical
  Physics} \textbf{2009}, {\it 130}, 094504.

\bibitem{Janco81}
B.~Jancovici, {\it Phys. Rev. Lett.} \textbf{1981}, {\it 46}, 386--388.

\bibitem{Alastuey81}
A.~Alastuey, B.~Jancovici, {\it J. Phys. France} \textbf{1981}, {\it 42},
  1--12.

\bibitem{Gaudin85}
M.~Gaudin, {\it J. Phys. France} \textbf{1985}, {\it 46}, 1027--1042.

\bibitem{Cornu87}
F.~Cornu, B.~Jancovici, {\it J. Stat. Phys.} \textbf{1987}, {\it 49}, 33--56.

\bibitem{Samaj2000}
L.~\ifmmode~\check{S}\else \v{S}\fi{}amaj, I.~Trav\v{e}nec, {\it J. Stat.
  Phys.} \textbf{2000}, {\it 101}, 713--730.

\bibitem{TellezTrizac15}
G.~T\'ellez, E.~Trizac, {\it Phys. Rev. E} \textbf{2015}, {\it 92}, 042134.

\bibitem{VarelaTT17}
L.~Varela, G.~T\'ellez, E.~Trizac, {\it Phys. Rev. E} \textbf{2017}, {\it 95},
  022112.

\bibitem{VarelaTT21}
L.~Varela, G.~T\'ellez, E.~Trizac, {\it Phys. Rev. E} \textbf{2021}, {\it 103},
  042603.

\bibitem{VarelaATT21}
L.~Varela, S.~Andraus, E.~Trizac, G.~Téllez, {\it J. Phys.: Condensed Matter}
  \textbf{2021}, {\it 33}, 394001.

\bibitem{Strat57}
R.~L. {Stratonovich}, {\it Soviet Physics Doklady} \textbf{1957}, {\it 2}, 416.

\bibitem{Hub59}
J.~Hubbard, {\it Phys. Rev. Lett.} \textbf{1959}, {\it 3}, 77--78.

\bibitem{ZZ79}
A.~B. Zamolodchikov, Al.~B. Zamolodchikov, {\it Annals of Physics}
  \textbf{1979}, {\it 120}, 253--291.

\bibitem{DV91}
C.~Destri, H.~J. {De Vega}, {\it Nuclear Physics B} \textbf{1991}, {\it 358}
  (1), 251--294.

\bibitem{Z95}
Al.~B. Zamolodchikov, {\it International Journal of Modern Physics A}
  \textbf{1995}, {\it 10} (08), 1125--1150.

\bibitem{LZ97}
S.~Lukyanov, Al. Zamolodchikov, {\it Nuclear Physics B} \textbf{1997}, {\it
  493}, 571--587.

\bibitem{L97}
S.~Lukyanov, {\it Physics Letters B} \textbf{1997}, {\it 408} (1), 192--200.

\bibitem{Smirnov1992}
F.~A. Smirnov, {\it {Form Factors in Completely Integrable Models of Quantum
  Field Theory}}, World Scientific, Singapore, \textbf{1992}.

\bibitem{FLZZ98}
V.~Fateev, S.~Lukyanov, A.~Zamolodchikov, Al. Zamolodchikov, {\it Nuclear
  Physics B} \textbf{1998}, {\it 516} (3), 652--674.

\bibitem{BS01}
P.~Baseilhac, M.~Stanishkov, {\it Nuclear Physics B} \textbf{2001}, {\it 612}
  (3), 373 -- 390.

\bibitem{Samaj2003}
L.~{\v{S}}amaj, {\it Journal of Statistical Physics} \textbf{2003}, {\it 111}
  (1), 261--290.

\bibitem{Tellez2005}
G.~T\'ellez, {\it Journal of Statistical Mechanics: Theory and Experiment}
  \textbf{2005}, {\it 2005} (10), P10001--P10001.

\bibitem{VarelaT21}
L.~Varela, G.~Téllez, {\it Journal of Statistical Mechanics: Theory and
  Experiment} \textbf{2021}, {\it 2021} (8), 083206.

\bibitem{SJ02}
L.~\v{S}amaj, B.~Jancovici, {\it J. Stat. Phys.} \textbf{2002}, {\it 106},
  301--312.

\bibitem{S05anomalous}
L.~\v{S}amaj, {\it J. Stat. Phys.} \textbf{2005}, {\it 120}, 125--146.

\bibitem{S06core}
L.~\v{S}amaj, {\it J. Stat. Phys.} \textbf{2006}, {\it 124}, 1179--1206.

\bibitem{Tellez2006EPL}
G.~T\'ellez, {\it Europhys. Lett.} \textbf{2006}, {\it 76} (6), 1186--1192.

\bibitem{HansenViot85}
J.~P. Hansen, P.~Viot, {\it Journal of Statistical Physics} \textbf{1985}, {\it
  38} (5), 823--850.

\bibitem{Tellez07}
G.~T\'ellez, {\it J. Stat. Phys.} \textbf{2007}, {\it 126}, 281--298.

\bibitem{cement21}
A.~Goyal, I.~Palaia, K.~Ioannidou, F.-J. Ulm, H.~van Damme, Roland J.-M.
  Pellenq, E.~Trizac, E.~Del Gado, {\it Science Advances} \textbf{2021}, {\it
  7} (32), eabg5882.

\end{thebibliography}

\end{document}